\documentclass[aps,prb,twocolumn,10pt]{revtex4-2}
\usepackage[utf8]{inputenc}
\usepackage{libertine}
\usepackage[libertine]{newtxmath}
\usepackage{mathtools,graphicx,microtype,siunitx}
\usepackage[cal=boondoxo,frak=boondox]{mathalfa}
\usepackage[colorlinks=true,allcolors=blue]{hyperref}

\DeclarePairedDelimiter\ket{\lvert}{\rangle}

\DeclarePairedDelimiterX\braket[2]{\langle}{\rangle}{#1 \delimsize\vert #2}
\DeclareMathOperator\arctanh{arctanh}
\makeatletter
\g@addto@macro\@floatboxreset\centering
\makeatother

\begin{document}

\title{The Coulomb impurity in 2D materials with strong spin-orbit interaction}

\author{Yasha Gindikin and Vladimir A.\ Sablikov}
\affiliation{Kotelnikov Institute of Radio Engineering and Electronics, Russian Academy of Sciences, Fryazino, 141190, Russia}

\begin{abstract}
We show that the spin-orbit interaction (SOI) produced by the Coulomb fields of charged impurities provides an efficient mechanism for the bound states formation. The mechanism can be realized in 2D materials with sufficiently strong Rashba SOI provided that the impurity locally breaks the structure inversion symmetry in the direction normal to the layer.
\end{abstract}

\maketitle 

Impurities play an important role in studying new materials not only because in many cases they dramatically affect the fundamental properties of the latter (a good example is the failure of the conductance quantization in the edge states of topological insulators~\cite{PhysRevLett.123.047701,PhysRevLett.122.016601}), but mainly because it is the electronic structure of the impurity states that the non-trivial material properties are manifested in most strikingly. 
Take for instance the phenomenon of falling to the center and the presence of a critical charge of impurities in graphene~\cite{Wang734,PhysRevB.76.233402}, the nontrivial electronic structure of impurities in 3D and 2D topological insulators~\cite{PhysRevB.81.233405,PhysRevB.85.121103,PhysRevB.91.075412}, numerous nontrivial manifestations of impurities in the properties of dichalcogenides~\cite{lin2016defect}, and so on.

In recent years, a great deal of attention was paid to materials with strong spin-orbit interaction (SOI)~\cite{manchon2015new,bihlmayer2015focus}. Yet, little is known about the electronic impurity states specific to these materials, although the scattering processes due to the SOI created by the charged impurities were widely studied in systems with strong Rashba effect~\cite{PhysRevLett.95.166605,PhysRevLett.106.126601,PhysRevB.93.085418} since the spin-dependent scattering owing to the SOI is the primary mechanism behind the acclaimed extrinsic spin-Hall effect~\cite{RevModPhys.87.1213}. However, recent studies have shown that in materials with strong Rashba SOI the electron energy arising due to the SOI created by the Coulomb field of a point charge can be comparable to the Coulomb energy~\cite{2019arXiv190506340G}, and for certain orientations of the spin and momentum of the electron this component of its interaction with the charge is attractive. Thus, this kind of SOI can not only significantly change the bound states in the Coulomb potential, but also lead to the formation of new states.

To explore this nontrivial possibility of bound states formation it is important to take into account the fact that the strong SOI is associated with the hybridization of basic Bloch states with different spin configurations; therefore, the study should be based on the multi-band model that describes the strong SOI\@. In this paper we use a 4-band model well justified for a wide class of materials with a strong SOI to show that the SOI created by the electric field of an impurity can be an effective mechanism for the formation of bound states with a high binding energy and specific spin structure. The bound states of a new type arise because the SOI leads to the effective attraction of the electron to the impurity charge of any sign.

Previously such attraction mechanism was studied for the pair electron-electron interaction in materials with strong SOI~\cite{2019arXiv190506340G} and was investigated within the conduction-band approximation~\cite{PhysRevB.98.115137,2018arXiv180410826G}. The attraction arises because the SOI lowers the energy of electrons in a certain spin configuration locked to their momenta. The pair SOI, being proportional to the Coulomb electric field of the charge the electron interacts with, grows like $\sim r^{-2}$ when their mutual distance $r \to 0$ goes to zero, whereas the Coulomb potential grows like $\sim r^{-1}$, which means that the attraction due to the SOI prevails over the Coulomb repulsion at a short distance. This leads to the formation of a bound electron pair. However, the solution loses stability because of the strong electric field divergence, which results in the wave-function collapse. In other words, there appears the ``fall to the center''. To regularize the solution it is necessary to go beyond the conduction-band approximation in order to take the SOI into account without the low-energy expansion. This problem is dealt with in the present paper for the case of a charged impurity.

Our study is based of the Bernevig–Hughes– Zhang (BHZ) model~\cite{Bernevig1757}, which is well substantiated and widely used to describe materials with a strong SOI and band structure formed due to the $sp^3$ hybridization. The model is built within the frame of $kp$ approximation and therefore well suited for the detailed study of the bound states. The BHZ model describes both trivial semiconductors with strong SOI and the topological phase. Being purely two-dimensional, this model describes correctly the effects of the in-plane electric field, and in particular the Rashba SOI produced by this field. In this model the impurity charge can be regarded as the source of the external field, however we should keep in mind that in a realistic system the charge is not located strictly in the center of the 2D layer but instead can be situated at any point within the layer width. This breaks the inversion symmetry with regards to the $z$-direction normal to the layer and hence the SOI appears defined by the normal component of the electric field $\mathcal{E}_{z}$. Our estimates show that the normal component of the electric field in the layer can become large enough to substantially alter the electron energy via the associated Rashba SOI\@. As a result, it is this component of the SOI that plays a decisive role in the formation of new bound states.

The generalization of the original BHZ model taking into account the normal electric field was done in Ref.~\cite{Rothe_2010}. We make use of this generalized model considering the impurity as the source of the external field, the two-dimensional potential and normal electric field component $\mathcal{E}_{z}$ of which are averaged across the layer thickness.

The four-band (taking the spin into account) Hamiltonian of the BHZ model is written in the basis

$(\ket{E+},\ket{H+},\ket{E-},\ket{H-})$, where $\ket{E \pm}$ and $\ket{H \pm}$ are the basis states of the electron-like and hole-like bands with angular momenta of $\pm 1/2$ and $\pm 3/2$, respectively:
\begin{equation}
    \label{ham}
    \hat{H} = 
    \begin{bmatrix}
        M(k)      & A k_+ & -i \xi e \mathcal{E}_{z} k_- &     0\\
        A k_-     & -M(k) & 0          &     0\\
        i \xi e \mathcal{E}_{z} k_+ & 0     & M(k)       & -A k_-\\
        0         &0      & -A k_+     & -M(k)
    \end{bmatrix} + U(r)I\,,
\end{equation}
with $k$ the momentum, $k_{\pm} = k_x \pm i k_y$, $M(k) = M - B k^2$, and $M$, $B$, $A$ being the model parameters. For simplicity, we consider the model symmetric with respect to electron-like and hole-like bands in the absence of the Rashba SOI\@. The topological phase is realized when $M B >0$, the trivial phase — when $M B <0$. Then, 
\begin{equation}
    \mathcal{E}_{z} = \frac{Z e}{\epsilon d} \left( \frac{1}{\sqrt{r^2 + \delta^2}} - \frac{1}{\sqrt{r^2 + {(d-\delta)}^2}} \right)\,,
\end{equation}
and
\begin{equation}
    U(r) = - \frac{Z e^2}{\epsilon d} \left( \arctanh \frac{d-\delta}{\sqrt{r^2 + {(d-\delta)}^2}} + \arctanh \frac{\delta}{\sqrt{r^2 + \delta^2}}\right)\,,
\end{equation}
with $d$ being the layer thickness, $Z$ the impurity charge in $e$ units, $\delta$ the impurity position as measured from the edge of the layer, $\epsilon$ the dielectric constant. To maintain the hermiticity of the Hamiltonian for the non-uniform normal field, the $\mathcal{E}_{z} k_{\pm}$ terms in Eq.~\eqref{ham} should be replaced by the anticommutator $\frac{1}{2}(k_{\pm}\mathcal{E}_{z} + \mathcal{E}_{z} k_{\pm})$. 

In Eq.~\eqref{ham}, $\xi$ is the parameter of the Rashba SOI that, clearly, couples only the electron-like states. Therefore the Rashba SOI affects the states in the conduction and valence bands in a non-symmetric way. Hence the attraction mechanism we are studying is non-symmetric with regards to the sign of the impurity charge $Z$ as well as the inversion of the electron-like and hole-like bands. For this reason the bound states are different in the trivial and topological phases.

The eigenfunctions of the Hamiltonian of Eq.~\eqref{ham} are
\begin{equation}
    \Psi = e^{i l \varphi} {(\psi_{1}(r), i \psi_{2}(r) e^{-i\varphi}, \psi_{3}(r) e^{i\varphi}, i \psi_{4}(r) e^{2i\varphi})}^{\intercal}\,,
\end{equation}
with integer $l$ being the angular quantum number.

The problem is treated numerically because the analytic approach is fraught with substantial difficulties. The system spectrum was found by the Petrov-Galerkin finite elements method~\cite{zienkiewicz} using the Arnoldi eigenvalue solver~\cite{golubvanloan}. 

We choose the system parameters close to those found in e.g.\ $\mathrm{HgTe/CdTe}$ quantum wells:  $ M = \SI{0.01}{eV}$, $A = \SI{5}{\angstrom eV}$, $B = -\SI{50}{\angstrom^2 eV}$, $\epsilon = 20$, $d = \SI{50}{\angstrom}$ and $\delta = \SI{3}{\angstrom}$. 

\begin{figure}[htb]
    \includegraphics[width=0.9\linewidth]{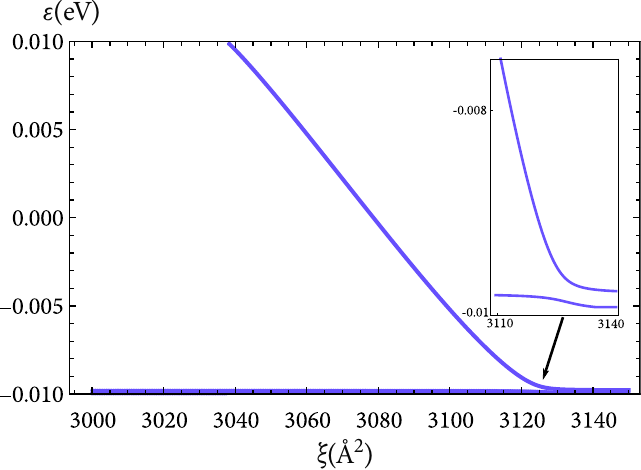}
    \caption{\label{trivial} The energy eigenvalues for $l=0$ in the gap as a function of $\xi$. The trivial phase, for negatively charged impurity with $Z=-2$. }
\end{figure} 

First consider the case of a negatively charged impurity in the trivial phase. The results of the calculations are shown in Fig.~\ref{trivial}. At small Rashba SOI, there exist only shallow hydrogen-like levels close to the valence band, the spectrum of which, hardly resolvable in the figure, was not investigated in detail as it is not of immediate interest. But for a critical value of $\xi$ there appears a new bound state, with the binding energy measured from the conduction band bottom rapidly increasing with $\xi$. As the term approaches the valence band top, there appears the anti-crossing between the term and the shallow hydrogen-like levels, as shown in the inset. 

\begin{figure}[htb]
    \includegraphics[width=0.9\linewidth]{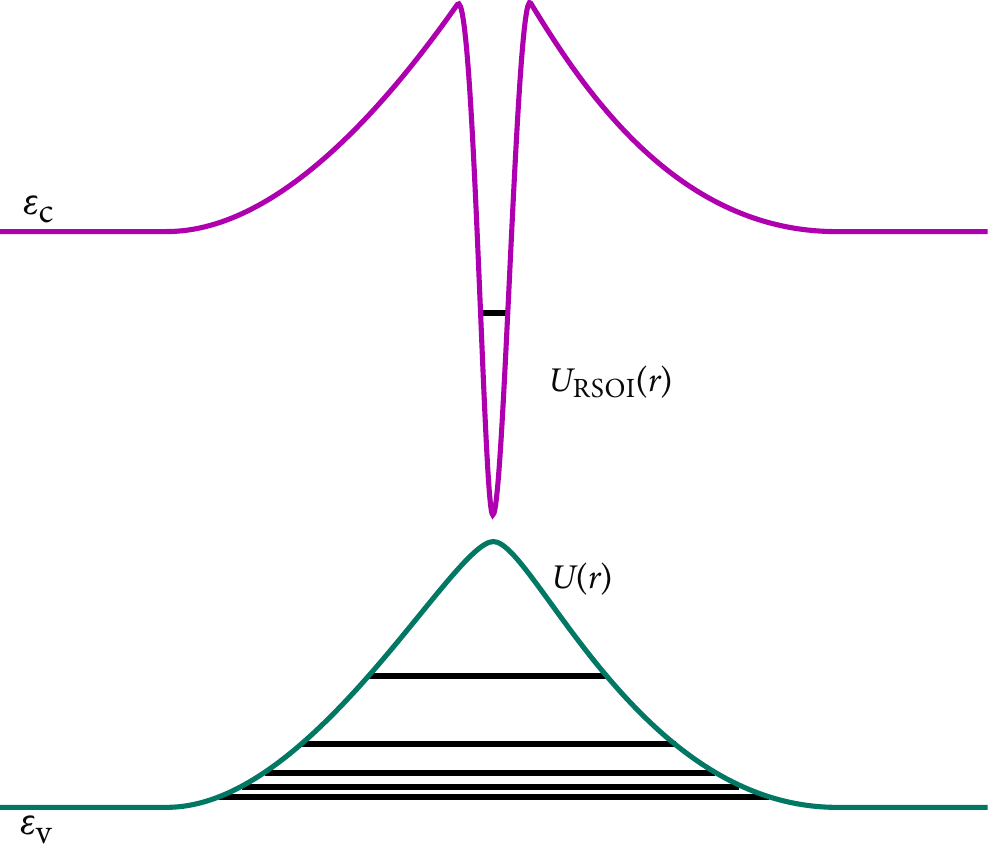}
    \caption{\label{pot} The schematic view of the distortion of the conduction band $\varepsilon_{\mathrm{c}}$ and valence band $\varepsilon_{\mathrm{v}}$ edges produced by the Coulomb potential of the negatively charged impurity $U(\mathbf{r})$ and the effective potential of the Rashba SOI $U_{\mathrm{RSOI}}(\mathbf{r})$.}
\end{figure}

This result clearly points to the following mechanism of the spectrum formation. On a qualitative level, let us assume that the impurity potential is smooth enough to create a spatially variating distortion of the band edges. Keep in mind that in addition to the Coulomb potential acting on both bands, there exist an effective potential of the Rashba SOI acting only on the electron states that form the conduction band in the trivial phase. Coulomb potential shifts the edges of both bands upwards, as it is shown in Fig.~\ref{pot}, whereas the Rashba SOI shifts the conduction band edge downwards. At sufficiently large $\xi$, a quantum well appears within the profile of the conduction band, the well width being small relative to the scale of the Coulomb potential.

The Coulomb potential creates the shallow states close to the valence band top, whereas the SOI potential, if larger then some critical value, creates a strongly localized state close to the conduction band bottom. The binding energy grows (i.e.\ the state becomes deeper) with increasing $\xi$. Generally speaking, the narrow well can contain multiple bound states with vastly different energies. This indeed follows from the numerical calculations that we do not include here. However such states are not interesting since their realization requires a too large Rashba SOI parameter or impurity charge.

The results shown in Fig.~\ref{trivial} as an illustration were obtained for a negatively charged impurity with a charge of $-2e$. Of course, the proposed mechanism for the bound state formation due to the SOI works for a charge of any magnitude and sign. With changing the magnitude of the charge the critical value of the parameter $\xi$ varies roughly as $1/|Z|$, and the binding energy of the states caused by SOI increases with increasing $|Z|$. The magnitude of the effective charge of an impurity in real materials depends on its electronic structure and bonds with the host material, and $Z=1\div 2$ seems to be a reasonable estimate.

\begin{figure}[htb]
    \includegraphics[width=0.9\linewidth]{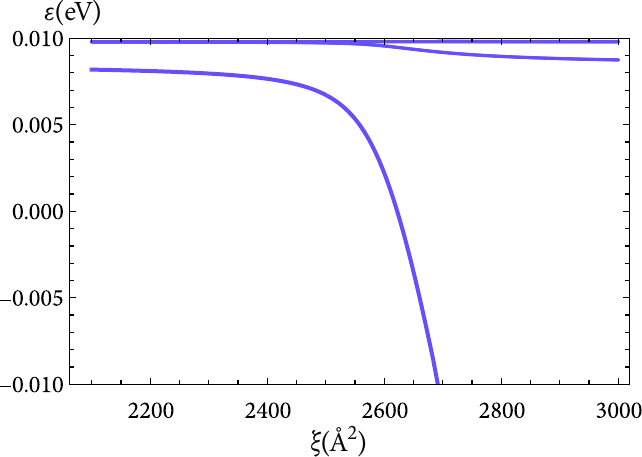}
    \caption{\label{at_trivial} The energy eigenvalues for $l=0$ in the gap as a function of $\xi$. The trivial phase, for positively charged impurity with $Z=+2$.}
\end{figure} 

Upon the change of the $Z$ sign, the spectrum is modified as follows, see Fig.~\ref{at_trivial}. The shallow hydrogen-like levels are formed in the vicinity of the conduction band bottom. At strong SOI, the lowest level rapidly drops down and for sufficiently large $\xi$ delves into the valence band. This spectrum can be easily understood by considering that the Coulomb potential of the positive impurity creates a potential well in the profile of the conduction band bottom, while the Rashba SOI deepens this well strongly in the vicinity of the Coulomb center.

Similar results are obtained also for the topological phase. The main qualitative difference of this case is that, due to the inversion of the electron and hole bands, the effective quantum well, which the SOI creates for a new kind of bound states, is formed relative to the valence band. In accordance with this, the spectrum of bound states and its dependence on the parameter $\xi$ also change. Otherwise, the picture and quantitative estimates are similar to those for the trivial phase.

To conclude, we have shown that the SOI produced by the Coulomb electric fields of the charged impurities gives rise to a new kind of bound states specific to materials with strong Rashba effect. The mechanism of their formation is due to the electron attraction to the charge of any sign, which appears thanks to the SOI created by the normal component of the Coulomb electric field of the impurity charge. It is important that the impurity locally breaks the structure inversion symmetry in the direction normal to the layer. Here we considered an impurity located asymmetrically in the layer. But this is not a unique possibility. Suitable structure inversion asymmetry can be created locally by the image charges induced by an impurity on a closely located gate.

On the contrary the possible structure inversion asymmetry in the bulk away from the impurity is not principally important, although it affects the bound states under investigation as well as the traditional hydrogen-like states in the Coulomb potential. This effect was explored for the topological phase previously~\cite{doi:10.1063/1.5049717} and is neglected here.

The proposed mechanism for the bound state formation is quite general and can be realized not only in materials described by the BHZ model with point-like defects. For materials like $\mathrm{HgTe/CdTe}$ we estimate the critical $\xi$ value to be $\SI{2.5e3}{\angstrom^2}$, which is twice as large as typical values known for such structures. In other materials with larger SOI, like $\mathrm{BiTeI}$, $\mathrm{Bi_2Se_3}$, $\mathrm{BiSb/AlN}$~\cite{ishizaka2011giant,PhysRevLett.107.096802,PhysRevB.95.165444,doi:10.1063/1.5074087}, and for other structure defects with larger charges~\cite{Wang734}, and in artificially created structures using e.g.\ the probe microscopy, such mechanism for the bound state formation can be quite realistic, but its investigation and the quantitative estimates require the model approaches to be different from those used here.

This work was carried out in the framework of the state task for IRE RAS and partially was supported by the Russian Foundation for Basic Research, project No. 20--02--0126.

\bibliography{paper}

\end{document}